\documentclass[aps,reprint]{revtex4-1}
\usepackage[]{graphicx}
\usepackage{color}
\newcommand{\vk}{\mbox{\boldmath $ k $}}

\newcommand{\vB}{\mbox{\boldmath $ B $}}

\newcommand{\vF}{\mbox{\boldmath $ F $}}
\newcommand{\vJ}{\mbox{\boldmath $ J $}}
\newcommand{\vf}{\mbox{\boldmath $ f $}}

\newcommand{\rr}{\mbox{\boldmath $ r $}}
\newcommand{\vU}{\mbox{\boldmath $ U $}}

\newcommand{\vomega}{\mbox{\boldmath $ \omega $}}
\newcommand{\btimes}{\mbox{\boldmath $ \times $}}

\begin{document}

%\preprint{}

\title{Hall-MHD small-scale dynamos} %Title of paper



\author{Daniel O. G\'omez}
\email[]{dgomez@df.uba.ar}
\homepage[]{http://astro.df.uba.ar}
%\thanks{}
\altaffiliation{also at Instituto de Astronom\'\i a y F\'\i sica del Espacio, C.C.~67 Suc.~28, 1428 Buenos Aires, Argentina.}
\author{Pablo D. Mininni}
%\thanks{}
\altaffiliation{also at IFIBA (Instituto de F\'\i sica de Buenos Aires), CONICET, 1428 Buenos Aires, Argentina; and  at National Center for Atmospheric Research, Boulder, CO 80307, USA. The National Center for Atmospheric Research is sponsored by the National Science Foundation.}
\author{Pablo Dmitruk}
%\thanks{}
\altaffiliation{also at IFIBA (Instituto de F\'\i sica de Buenos Aires), CONICET, 1428 Buenos Aires, Argentina.}
\affiliation{Departamento de F\'\i sica, Facultad de Ciencias Exactas 
    y Naturales, Universidad de Buenos Aires and CONICET, Ciudad 
    Universitaria, 1428 Buenos Aires, Argentina.}
%\noaffiliation is required (may also be used with the \author command).

\date{\today}

\begin{abstract}
Much of the progress in our understanding of dynamo mechanisms has been made within the theoretical framework of magnetohydrodynamics (MHD). However, for sufficiently diffuse media, the Hall effect eventually becomes non-negligible. We present results from three dimensional simulations of the Hall-MHD equations subjected to random non-helical forcing. We study the role of the Hall effect in the dynamo efficiency for different values of the Hall parameter, using a pseudospectral code to achieve exponentially fast convergence. We also study energy transfer rates among spatial scales to determine the relative importance of the various nonlinear effects in the dynamo process and in the energy cascade. The Hall effect produces a reduction of the direct energy cascade at scales larger than the Hall scale, and therefore leads to smaller energy dissipation rates. Finally, we present results stemming from simulations at large magnetic Prandtl numbers, which is the relevant regime in hot and diffuse media such a the interstellar medium.
\end{abstract}
%\pacs{47.27.-i Turbulent flows, 95.30.Qd Magnetohydrodynamics and plasmas, 95.50.Bh Interplanetary magnetic fields}

\maketitle %\maketitle must follow title, authors, abstract and \pacs

\section{Introduction}\label{sec:intro}
%\subsection{}
%\subsubsection{}
The generation of magnetic fields by dynamo activity plays an important role in a wide range of astrophysical objects, ranging from stars to clusters of galaxies. The gas in these objects is characterized by turbulent flows, as shown for instance by scintillation observations of the interstellar medium \cite{Span01, Mint96}, or from pressure maps in galaxy clusters \cite{Schue04}. Mechanisms able to generate magnetic fields by dynamo action are often classified as large- and small-scale dynamos, depending on the correlation length of the induced magnetic field. In this context, large and small are referred to the energy containing scale of the turbulent hydrodynamic flow. This classification is not rigid, as in many astrophysical objects both dynamos may be at work, but it gives a useful framework considering the limitations in the scale separation that can be achieved in numerical simulations. Also, the physical properties of the flows that can give rise to one or the other are somewhat different.

Helical flows have proved efficient in generating large-scale dynamos, i.e., on scales larger than the energy-containing eddies of the flow \cite{Pouq76,Meneg81,Brand01,Gomez05}. It is now known that large-scale dynamo action can also be produced by anisotropic and inhomogeneous flows (e.g., flows with a large scale shear).  On the other hand, non-helical flows can be instrumental in generating small-scale dynamos \cite{Kazantsev68}, i.e., on sizes smaller than those of the energy-containing eddies \cite{Scheko01,Scheko04a,Haugen04}. In recent years, the study of small-scale dynamos with magnetic Prandtl number $Pm = \nu /\eta $ (the ratio between the viscosity and the magnetic diffusivity of the plasma) different from unity has received special attention \cite{Scheko04b,Mininnietal05}, both for $Pm \gg 1$ \cite{Scheko02} and for $Pm \ll 1$ \cite{Ponty05,Iska07}.  Motivations to study these regimes include recent experiments of dynamo action using liquid sodium \cite{Monchaux07}, as well as the fact that many astrophysical plasmas are characterized by magnetic Prandtl numbers different from unity. For instance, the magnetic Prandtl number is much smaller than one in the solar convective region, and it is typically much larger than one in the interplanetary medium and also in the interstellar medium (ISM).

For sufficiently low-density media such as the one that pervades the ISM, kinetic effects such as the Hall effect or ambipolar diffusion might also become relevant \cite{Sano02}. The potential relevance of ambipolar diffusion in astrophysical dynamos was studied in Refs.~\cite{Branden00,Zweibel02}. The relevance of the Hall effect has been recognized in various astrophysical applications \cite{Balbus01,Sano02,Mininni02}, space plasmas \cite{Deng01,Oieroset01,Mozer02}, and also laboratory plasmas \cite{Yamada97,Mirnov03,Ji08}. The role of the Hall effect on large-scale dynamos subjected to helical forcing has also been addressed in the literature \cite{Mininni03a,Mininni03b}. Less attention has received the impact of kinetic effects on the small-scale dynamo. A theoretical model of the kinematic small-scale dynamo with Hall effect was presented in \cite{Kleeorin94}, but to the best of our knowledge no numerical studies of the non-linear and saturated regime were considered in the literature.

In this paper, we present results from three dimensional simulations of the Hall-MHD equations subjected to random non-helical forcing. The main aim is to study the role of the Hall effect in the small-scale dynamo efficiency for different values of the Hall parameter. As a result of the study, we also discuss the impact of the Hall effect on the dynamo saturation values, and on magnetic and total dissipation rates. The structure of the paper is as follows.  A brief introduction to the theoretical framework known as Hall-MHD is presented in Sect.~\ref{sec:hmhd}. The role of the Hall effect in the efficiency of the dynamo is shown in Sect.~\ref{sec:lin}. In Sect.~\ref{sec:spec} we characterize the stationary regime that is attained when the dynamo process saturates, showing the corresponding energy power spectra. The energy transfer rates participating in the nonlinear energy cascade are displayed in Sect.~\ref{sec:trans}. In Sect.~\ref{sec:Pm}, we explore the regime of large magnetic Prandtl number (i.e., when the viscous dissipation scale is larger than the resistive dissipation scale) which, as mentioned, is particularly relevant in diffuse media such as the ISM. Finally, the conclusions are summarized in Sect.~\ref{sec:conclu}.

\section{Hall-MHD equations}\label{sec:hmhd}
For the sake of simplicity, we consider incompressible flows, although compressible effects may be relevant, e.g., in the ISM for the formation of structures \cite{Passot95}. Incompressible Hall-MHD is described by the modified induction equation (i.e., with the addition of the Hall current) and the equation of motion (the Navier-Stokes equation),
\begin{eqnarray}
\frac{\partial\vB}{\partial t} & = & \nabla\btimes\left[\left( \vU
     - \epsilon\nabla\btimes\vB\right)\btimes\vB\right] 
     + \eta \nabla^2 \vB \label{HallMHD} \\
\frac{\partial \vU}{\partial t} & = & - \left( \vU \cdot \nabla \right) \vU + \left( \vB \cdot \nabla \right) \vB -
 \nabla \left( P + \frac{B^2}{2} \right) + \nonumber \\ 
\mbox{} && \vF + \nu \nabla^2 \vU , 
\label{NS}
\end{eqnarray}
where $\vF$ denotes a solenoidal and non-helical external force, which is delta-correlated in time. The velocity $\vU$ and the magnetic field $\vB$ are expressed in units of a characteristic speed $ U_0 =\sqrt{\left<U^2\right>}$; $ \eta $ is the magnetic diffusivity, and $ \nu $ is the kinematic viscosity. The parameter $\epsilon $ measures the relative strength of the Hall effect and can be written as 
\begin{equation}
\epsilon = \frac{c}{\omega_{pi} L_0}\ \frac{U_A}{U_0} ,
\label{eps}
\end{equation}
where $ L_0 $ is a characteristic length scale, $U_A=\sqrt{\left<B^2\right>/4\pi n m_i}$ is the Alfven speed, and $w_{pi}=\sqrt{4\pi e^2 n/m_i}$ is the ion plasma frequency ($e$: electron charge, $n = n_e = n_i$: particle, electron, and ion density respectively, and $m_i$: ion mass). Hereafter, we adopt $U_0 = U_A$ as our typical velocity, thus rendering the Hall parameter simply as $\epsilon = c/(\omega_{pi} L_0)$, i.e., a dimensionless version of the ion skin depth.

These equations are complemented by the solenoidal conditions for both vector fields, i.e.,
\begin{equation}
\nabla\cdot\vB = 0 = \nabla\cdot\vU\ .
\label{div}
\end{equation}

From a theoretical point of view, Hall-MHD corresponds to a two-fluid description of a fully ionized plasma: a positively charged ion species of mass $m_i$ moving with the velocity field $\vU (\rr , t) $, and negatively charged massless electrons with the velocity
\begin{equation}
\vU_e = \vU - \epsilon\nabla\btimes\vB .
\label{ue}
\end{equation}
Therefore, from Eqs. (\ref{HallMHD})-(\ref{ue}) we obtain that in the ideal limit (i.e. $\eta\rightarrow 0$), the magnetic field is frozen to the electron flow. As a result, advection, stretching, and folding of magnetic field lines (mechanisms relevant for dynamo action) are performed by the electron flow rather than by the bulk flow, resulting in potential modifications to magnetic field generation when the Hall effect is not negligible.

\section{Linear and nonlinear dynamo efficiencies}\label{sec:lin}

We performed simulations of the Hall-MHD equations with a spatial resolution of $256^3$ gridpoints, using a pseudospectral code \cite{Mininni05}. The linear size of our numerical box is $2\pi L_0$ (with $L_0$ a unit length), and periodic boundary conditions in the three cartesian directions are assumed. We apply the $2/3$ dealiasing rule, and therefore the maximum wavenumber resolved by the code is $k_{max}=256/3\approx 85$. We first consider simulations with magnetic Prandtl number equal to unity (i.e., $Pm=1$). The coefficients of viscosity and resistivity in these simulations are set to $\nu = \eta = 2\times 10^{-3}$, which ensure that the dissipation scales are well resolved, i.e., at all times the dissipation wavenumbers $k_\nu =(\left<\omega^2\right>/\nu^2)^{1/4}$ and $k_\eta =(\left<J^2\right>/\eta^2)^{1/4}$ remain smaller than $k_{max}$ (here, $\vomega = \nabla \times \vU$ is the vorticity, and $\vJ = \nabla \times \vB$ is the current density). To evolve the equations in time we use a fully explicit second order Runge-Kutta scheme. We note that for Hall-MHD, and for velocity and magnetic fields of order unity, the Courant-Friedrichs-Levy (CFL) condition becomes $\Delta t \leq (\Delta x)^2/\epsilon$ (due to the dispersive nature of the whistler waves), which is more restrictive than the regular CFL condition $\Delta t\leq \Delta x$. As a result, the time step decreases quadratically with the spatial resolution, and also decreases linearly with the Hall parameter $\epsilon$. The Hall-MHD dynamo simulations are therefore computationally more expensive than the equivalent MHD runs, resulting in the modest spatial resolution considered here.

We first generate stationary hydrodynamic turbulence by integrating Eq. (\ref{NS}) subjected to random non-helical forcing (i.e., such that $\nabla\times\vF\ \perp\ \vF$) centered at wavenumbers $|\vk | \approx k_F = 3$ and delta-correlated in time. Once the kinetic energy reaches a stationary regime as a result of the balance between the power delivered by the external force and viscous dissipation, the hydrodynamic simulation is stopped. In a second stage, a random and small magnetic field is introduced at small scales, and the simulation is restarted with the full Hall-MHD equations (\ref{HallMHD})-(\ref{NS}). 

We performed simulations with different values of the Hall parameter $\epsilon$, including a purely MHD case corresponding to $\epsilon = 0$. Whenever $\epsilon \neq 0$, a new spatial scale is introduced (the Hall scale), which in the spectral domain is characterized by $k_\epsilon = 1/\epsilon$. In this paper, we consider the cases in which $k_\epsilon$ falls in between the macroscopic scale $k_F$ (set by the external driver) and the dissipation scales $k_\nu$ and $k_\eta$, which is the relevant scenario for astrophysical plasmas such as the interstellar medium. In such media, the Hall scale is several orders of magnitude smaller than the largest scales, and the Hall effect can be expected to be relevant only at the smallest dynamical scales. Note, however, that these arguments ought to be regarded as motivations. Although the ordering of typical length scales is the correct one, a realistic separation of scales is completely out of reach with present computing power.

\begin{figure}
{\includegraphics{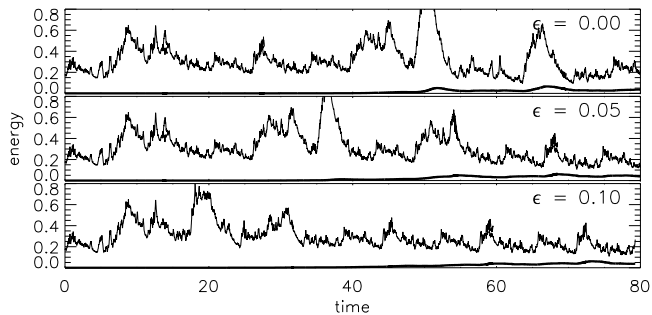}}
\caption{Kinetic (thin) and magnetic (thick) energies vs.~time for $\epsilon = 0$, $0.05$, and $0.10$ (from top to bottom).}\label{fig:ener}
\end{figure}

\begin{figure}
{\includegraphics{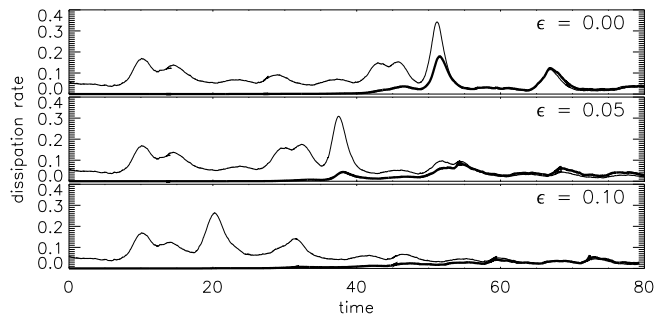}}
\caption{Kinetic (thin) and magnetic (thick) dissipation rates vs.~time for $\epsilon = 0$, $0.05$, and $0.10$ (from top to bottom).}\label{fig:diss}
\end{figure}

In Fig. \ref{fig:ener} we show the statistically stationary time series for kinetic energy (thin lines) for runs with different values of the Hall parameter $\epsilon$. The magnetic energy in these runs (thick line) is observed to rise until it saturates at values which remain a moderate fraction of the corresponding kinetic energy. The viscous (thin line) and resistive (thick) dissipation rates vs.~time are shown in Fig. \ref{fig:diss} for three runs with different values of the Hall parameter. In all these plots, the magnetic dissipation rate is observed to grow until it becomes fully comparable to the corresponding viscous dissipation rate (even when the kinetic energy is larger than the magnetic energy).

\begin{figure}
\includegraphics{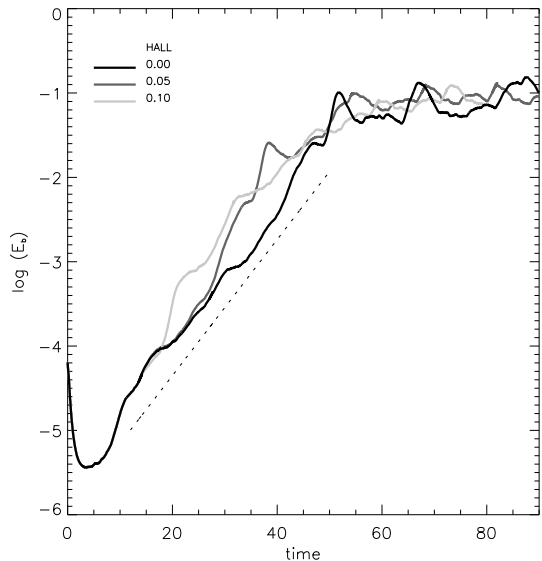}
\caption{Magnetic energy vs.~time, showing the exponential growth rate during the early linear dynamo regime.}\label{fig:b2}
\end{figure}

The exponentially fast growth of magnetic energy is shown using a lin-log scale in Fig. \ref{fig:b2} for the same three runs. Note that there is an initial stage where the magnetic field starts growing exponentially fast, regardless of the particular value of the Hall parameter $\epsilon$. During this early stage of the dynamo, the electron flow is still approximately equal to the ion flow, i.e. $\vU_e \approx \vU$ [see Eq. (\ref{ue})]. Keeping in mind that the growing magnetic field remains approximately frozen (note that this is strictly valid only in the limit $\eta\rightarrow 0$) to the electron velocity field, we can anticipate that at some point in time the electron and ion flows will start drifting appart from one another. Therefore, a second stage arises corresponding to a non-linear dynamo (although still ``kinematic,'' in the sense that the magnetic field does not affect the bulk velocity field), since the magnetic field is being advected by the electron flow which at that point becomes a function of the magnetic field itself. In Fig. \ref{fig:b2} we see that although the case $\epsilon = 0$ can be approximated by a linear growth rate (indicated by the dotted straight line) all the way up to the saturation level, we cannot do the same for the cases $\epsilon = 0.05$, and $0.10 $ since there is a break in the corresponding growth rates. This break occurs first for the case with larger Hall effect (i.e., $\epsilon = 0.10$), but the incremented slope is larger for the case $\epsilon = 0.05$. The fact that the dynamo efficiency improves up to a certain value of the Hall parameter and then starts decreasing, is reminiscent of similar results reported in Ref. \cite{Mininni05} for large-scale Hall-MHD dynamos. 

\begin{figure}
 \includegraphics{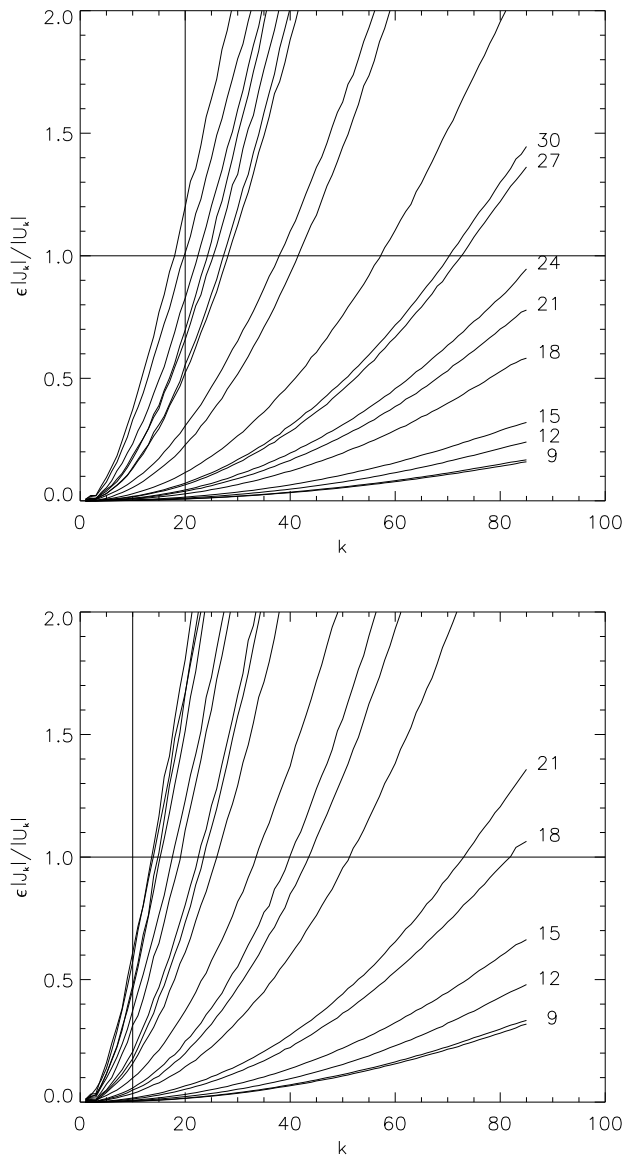}
\caption{Ratio between $\epsilon|J_k|$ and $|U_k|$ at different times (labeled). The top frame corresponds to the run with $\epsilon = 0.05$ and the bottom frame to $\epsilon = 0.10$. The vertical gray line in each frame corresponds to the Hall wavenumber $k_\epsilon = 1 / \epsilon $.}\label{fig:eju}
\end{figure}

To show the relative importance of the Hall term in the electron velocity field, in Fig.~\ref{fig:eju} we display the ratio between $\epsilon|J_k|$ and $|U_k|$ at different labeled times, where $|J_k|$ and $|U_k|$ are respectively the spectral intensities of the current density and of the velocity field at wavenumber $k$ [note that $\vU_e = \vU - \epsilon \vJ$ from Eq. (\ref{ue})]. The upper frame corresponds to the run with $\epsilon = 0.05$ and the lower frame to $\epsilon = 0.10$. The vertical gray line in each frame corresponds to the Hall scale $k_\epsilon = 1 / \epsilon $. In both cases, the Hall term becomes gradually non-negligible and eventually dominant at the largest wavenumbers of the system, i.e.,  $\epsilon|J_k| > |U_k|$ at $k > k_\epsilon$. For the case $\epsilon = 0.05$ (upper panel), the Hall term   $\epsilon|J_k|$ becomes comparable to $|U_k|$ at the largest wavenumbers by about $t\approx 24$, while a similar situation arises for $\epsilon = 0.10$ at $t\approx 18$ (lower panel). These values of time are remarkably consistent with those observed in Fig. \ref{fig:b2} for the departure from the linear regime in each of the runs.

In all these cases, there is a third and last stage, which corresponds to the saturation of the dynamo. We discuss the energy saturation levels in the next section.

\section{Energy spectra}\label{sec:spec}

\begin{figure}
\includegraphics{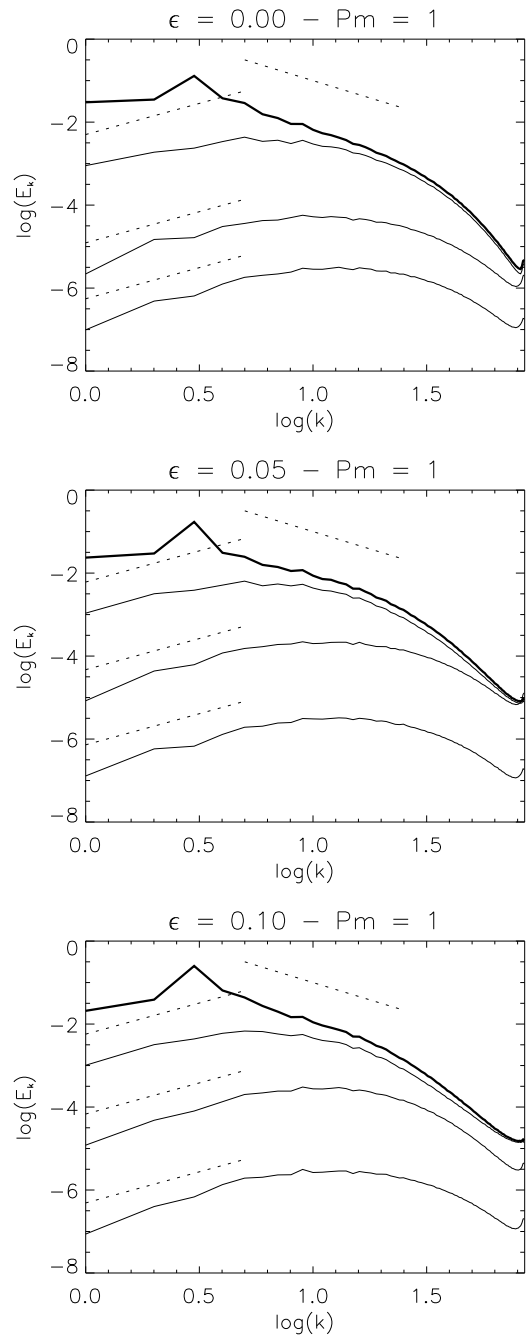}
\caption{Total energy spectrum (thick trace) at $t=72$ for different values of $\epsilon$ (labeled). In each frame magnetic energy spectra at $t = 18, 36, 72$  are also shown (corresponding to the thin lines from bottom to top). The Kolmogorov and Kazantsev spectra are overlaid (dotted trace) for reference.}\label{fig:sp}
\end{figure}

The distribution of kinetic and magnetic energy among spatial scales can be observed in the energy power spectra $E_k$ vs.~$k$ displayed in Fig. \ref{fig:sp} for three different values of the Hall parameter ($E_k$ is defined such that the total energy is $E = \int dk\ E_k$, and magnetic and kinetic energies are such that $E_b + E_u = E$). The thick line in all these runs corresponds to the total energy spectrum, which remains in a roughly statistically stationary regime. Note that for these three runs, the kinetic energy spectrum remains close to Kolmogorov, i.e., $E_k \propto k^{-5/3}$, in the energy inertial range. The thin lines display magnetic energy spectra at different times, showing the growth of magnetic energy as a whole at early times, and saturation of magnetic field growth at small scales at intermediate and late times. Note that the peak of the magnetic energy in all these cases remains at wavenumbers longer than the one where the external force operates (i.e., $k_F = 3$), as expected for a small-scale dynamo.

The Kazantsev slope $E_k \propto k^{3/2}$ \cite{Kazantsev68} provides a reasonable approximation at small wavenumbers for all these cases. Kazantsev's dynamo theory \cite{Kazantsev68} assumes a random velocity field with Gaussian statistics, which is spatially homogeneous and isotropic and $\delta$-correlated in time. Under these assumptions, the two-point magnetic field correlation function can be analytically computed (see Ref.~\cite{Kazantsev68}, also Ref.~\cite{Branden05}), and a $k^{3/2}$ power law is asymptotically expected for the magnetic energy spectrum at the low wavenumber end. Even though Kazantsev's model was devised for pure MHD (no Hall effect) and for large magnetic Prandtl numbers, our Hall-MHD simulations also reproduce an $E_k \propto k^{3/2}$ magnetic spectrum equally well. This is to be expected, since the Hall effect becomes negligible at the lowest wavenumbers (i.e., at $k \ll k_\epsilon = 1/\epsilon$). Kazantsev's spectrum has also been reported in simulations of small scale MHD dynamos at unity Prandtl numbers \cite{Haugen04}. The extension of Kazantsev's model to Hall-MHD in Ref.~\cite{Kleeorin94} also recovers this spectrum in the regime considered here.

In summary, a preliminary inspection of the magnetic energy power spectra at early times shows no noticeable differences between MHD and Hall-MHD. On the one hand this is not surprising, since the Hall effect is nonlinear in the magnetic field, and the magnetic energy at early times is much smaller than the kinetic energy at all scales. On the other hand, in what follows we show that this last conclusion is somewhat premature, since there are other aspects of these turbulent dynamos that clearly show the consequences of the Hall effect.

At saturation, the total magnetic energy reaches a sizeable fraction of the total kinetic energy, which can be estimated within $10 \%$ to $20 \%$. More specifically, after taking time averages between $t=60$ and $t=80$ (see Figs.~\ref{fig:diss} and \ref{fig:b2}), we obtain the energy ratios $E_b/E$ listed in Table \ref{tab:1}. Note that the saturation level of these small-scale dynamos, defined as the fraction of magnetic energy to total energy in the stationary regime, decreases with the Hall parameter. Therefore, although in the linear dynamo regime the growth rate increases with the Hall parameter $\epsilon$, the magnetic field reaches a smaller saturation level.

\begin{table}
\caption{\label{tab:1} Global results for runs with different values of the Hall parameter $\epsilon$. $E$ is the mean saturation level of the total energy, $E_b/E$ is the ratio of magnetic to total energy, $k_J$ is the average wavenumber for the current density distribution, $D$ is the total dissipation rate, and $D_b/D$ is the ratio of magnetic to total dissipation rate. }
\begin{ruledtabular}
\begin{tabular}{cccccccc}
 & $\epsilon$ & E & $E_b /E $ & $k_J$ & D & $D_b /D $ & \\
\hline
 & 0.00 & 0.37 & 0.14 & 23.1 & 0.13 & 0.48 & \\
 & 0.05 & 0.35 & 0.13 & 19.4 & 0.11 & 0.39 & \\
 & 0.10 & 0.33 & 0.13 & 17.4 & 0.10 & 0.34 & \\
\end{tabular}
\end{ruledtabular}
\end{table}

\begin{figure}
\includegraphics{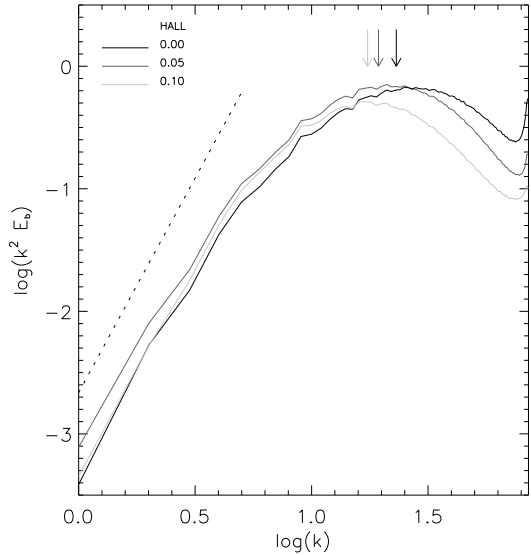}
\caption{Spectral distribution of current density, i.e., $k^2 E_b(k)$ vs.~$k$ for three different values of the Hall parameter (labeled). The dotted trace corresponds to the Kazantsev slope $k^{7/2}$. The arrows indicate the average wavenumber $k_J$ (see Eqn.~\ref{kJ}) for each distribution.}\label{fig:j2k}
\end{figure}

As mentioned in Sect.~\ref{sec:lin}, the dynamics of the largest wavenumbers in our simulations is controlled by viscosity and electric resistivity. Therefore, the dissipation of magnetic energy mostly takes place in current sheets with a thickness which can be expected to be close to the inverse of $k_\eta \approx 85$. On the other hand, the width and the length of these current sheets will vary from one to another \cite{Scheko04b,Servidio09}. We can obtain a statistical average of the dimensions of our magnetic dissipative structures by computing the power spectrum of the electric current density, which is simply $k^2E_b(k)$. In Fig. \ref{fig:j2k} we show time averaged (between $t = 60$ and $t = 80$) current density spectra for three different values of the Hall parameter (labeled). All of these spectra are compatible with a Kazantsev law at low wavenumbers. Note that the maximum of these spectra shift toward smaller wavenumbers as the Hall parameter increases. Since the peak of the spectrum can be associated to an average thickness of the current sheets, the above mentioned shift can be interpreted as the current sheets becoming relatively ``thicker'' as the Hall effect increases. This result is in agreement with previous experimental and numerical results suggesting that in Hall-MHD the thickness of the current sheets is given by the Hall scale  rather than by the Ohmic dissipative scale as in the MHD case (see \cite{Dmi06} for recent results in support of this interpretation). 

For each of these runs, we also compute the magnetic Taylor wavenumber $k_J$ given by

\begin{equation}
k_J^2 = \frac{\int dk\ k^2\ E_b(k)}{\int dk\ E_b(k)}
\label{kJ}
\end{equation}

\noindent which are indicated in Figure~\ref{fig:j2k} by arrows, and are observed to remain close, but somewhat to the left of the maximum for the corresponding power spectrum. The magnetic Taylor scale (i.e., the inverse of $k_J$) can be interpreted as the mean curvature of the magnetic field lines \cite{Scheko04b} and of the ensuing current sheets. The value of $k_J$ also moves towards smaller wavenumbers as the Hall scale is increased. The values of $k_J$ for each of these runs are listed in Table \ref{tab:1}. In Table \ref{tab:1} we also list the time averaged total dissipation rate $D = D_u + D_b$, clearly showing a progressive reduction as the Hall parameter is increased. The ratio of magnetic to total dissipation $D_b/D$ also reduces as $\epsilon$ increases, going from approximate equipartition in the MHD case to about $33 \%$ for $\epsilon = 0.10$, even though in all these simulations the relative content of magnetic energy $E_b/E$ is comparatively much smaller.

\section{Energy transfer rates}\label{sec:trans}

Interpretation of these results on the energy dissipation rate requires a detailed analysis of the transfer and conversion rate of energy among scales and between the velocity and magnetic fields, in order to identify the sources of small-scale dynamo action in MHD and in Hall-MHD turbulence. Equations (\ref{HallMHD})-(\ref{NS}) are known \cite{Turner86} to have three ideal invariants: energy, magnetic helicity and hybrid helicity. These are transferred between scales without losses by the non-linear terms in Eqs. (\ref{HallMHD})-(\ref{NS}). In this paper we focus our attention in the transfer and conversion of energy,
\begin{equation}
E = \frac{1}{2}\ \int\ d^3r\ (|\vU|^2+|\vB|^2) = \int dk\ E_k\  ,
\label{energy}
\end{equation}
since the non-helical dynamo does not generate helical magnetic fields. The dynamo process in this case is basically the conversion of mechanical energy into magnetic energy by induction, to sustain the magnetic fields against Ohmic dissipation. The nonlinear terms in Eqs.~(\ref{HallMHD})-(\ref{NS}) only redistribute energy (and the other ideal invariants) among different spatial scales, in such a way to guarantee the global invariance except for dissipative losses. To study the energy transfer in Fourier space, we follow the method of shell filters originally developed for MHD \cite{Verma04,Alexakis05}, which has more recently been extended for Hall-MHD \cite{Mininni07}. For a given vector field $\vf (\rr )$, we define  $\vf_\kappa (\rr ) $ to be the composition of all wavenumbers between $\kappa$ and $\kappa + 1$, i.e.,
\begin{equation}
\vf_\kappa (\rr ) = \sum_{|\vk|=\kappa}^{\kappa+1} \hat{\vf}(\vk)\ e^{i\vk\cdot\rr}
\label{fK}
\end{equation}
so that $\vf (\rr ) = \sum_{\kappa=0}^\infty  \vf_\kappa (\rr )$. From Eqs.~(\ref{HallMHD})-(\ref{NS}) we can derive detailed balance equations for the kinetic and magnetic energy in the $\kappa$-shell
\begin{widetext}
\begin{eqnarray}
\partial_t E_U(\kappa) & = & \int d^3r \bigg[ \sum_Q \Big[
    \overbrace{-\vU_\kappa\cdot(\vU\cdot\nabla )\vU_Q}^\textrm{Advection} +
    \overbrace{\vU_\kappa\cdot(\vB\cdot\nabla )\vB_Q}^\textrm{Lorentz} \Big] +
    \overbrace{\nu\vU\cdot\nabla^2\vU_\kappa}^\textrm{Dissipation} + 
    \overbrace{\vF\cdot\vU_\kappa}^\textrm{Injection} \bigg] \label{Eu} \\
\partial_t E_B(\kappa) & = & \int d^3r \bigg[ \sum_Q \Big[
    \underbrace{-\vB_\kappa\cdot(\vU\cdot\nabla )\vB_Q}_\textrm{Advection} +
    \underbrace{\vB_\kappa\cdot(\vB\cdot\nabla )\vU_Q}_\textrm{Induction} 
    \Big] +
    \underbrace{\eta\vB\cdot\nabla^2\vB_\kappa}_\textrm{Dissipation} + 
    \underbrace{\epsilon\vJ_\kappa\cdot(\vB\times\vJ_Q)}_\textrm{Hall}
    \bigg] \label{Eb}
\end{eqnarray}
\end{widetext}

\begin{figure}
\centerline{
\includegraphics{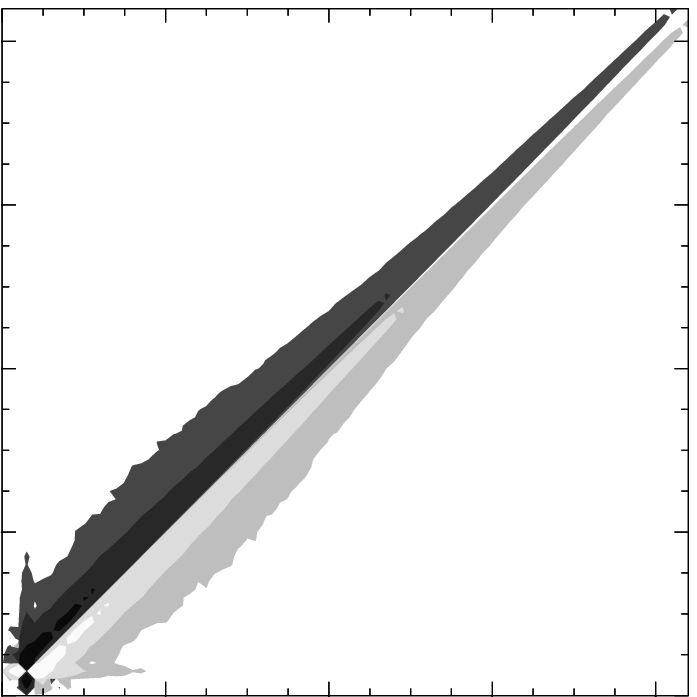}}
\centerline{
\includegraphics{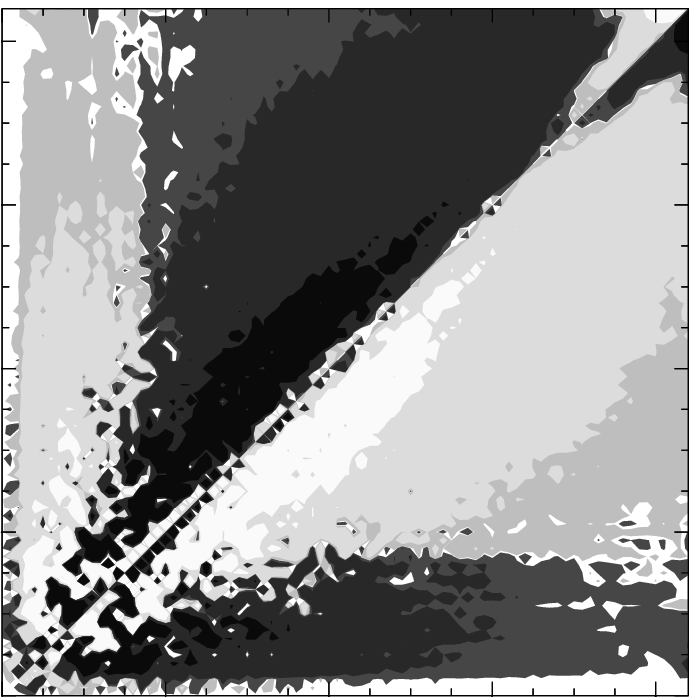}}
\caption{Energy transfer rate contour plots on the $(\kappa , Q )$ plane ($\kappa$ corresponds to the horizontal axis) for the run with $\epsilon = 0.05$ at the stationary regime. Both $\kappa$ and $Q$ run from zero to $k_{max} = 85$. Light-gray filled contours correspond to levels at [0.001,0.010,0.100] of the maximum positive value, while dark-gray contours display the same levels at negative energy transfer rates. The top frame shows the total transfer rate, with a peak value of $0.36$, and the bottom frame shows the Hall transfer rate, with a peak value of $4\times 10^{-4}$. }\label{fig:trans}
\end{figure}

All cubic terms in Eqs. (\ref{Eu})-(\ref{Eb}) can be interpreted as energy transfer from the third field in the $Q$-shell to the first field in its $\kappa$-shell, and associated to different physical processes as indicated in the equations. For instance, Fig. \ref{fig:trans}a shows the total shell-to-shell energy transfer rate [i.e., the sum of all cubic terms in Eqs.~(\ref{Eu})-(\ref{Eb})] in the steady state of the run with $\epsilon=0.05$. Light-gray (dark-gray) contours correspond to positive (negative) energy transfer regions on the $(\kappa , Q)$ plane, located at fractions of $[0.001,0.010,0.100]$ of the maximum positive (minimum negative) value. The first thing to notice is that the integral of this function on the $(\kappa , Q)$ plane amounts to zero, which is expected to be the case for any conserved quantity in the ideal limit. The second aspect to notice, is the high degree of concentration around the region $Q\approx \kappa$, which is indicative of the mostly local nature of the direct cascade of total energy in Hall-MHD, just as for MHD turbulence (for a detailed study of local and non-local contributions to the cascade in MHD and Hall-MHD, see Refs.~\cite{Alexakis05,Mininni07}). In Fig.~\ref{fig:trans}b we show the same kind of plot for just the Hall cubic term [see Eq.~(\ref{Eb})]. The Hall transfer is non-local, although it is important to emphasize that this term is much smaller than the total transfer rate. The maximum value for the Hall transfer rate is only $4 \times 10^{-4}$, while the total transfer rate peaks at about $0.36$.

The elongated dark spot in the lower part of Fig. \ref{fig:trans}b, more specifically in the region $Q < k_\epsilon = 20$, indicates that energy is transfered backwards by the Hall term from small scales to scales larger than the Hall scale. On the other hand, the light spot below the diagonal (with the dark triangle above the diagonal) for $\kappa , Q > k_\epsilon = 20$, indicates that at scales smaller than the Hall scale, the Hall term contributes to the direct energy cascade increasing the transfer rate towards smaller scales.

We can also compute the energy flux at the wavenumber $|\vk| = k$ by simply performing
\begin{widetext}
\begin{equation}
\Pi (k) = \sum_{\kappa=0}^k \sum_Q \int d^3r  \Big[-\vU_\kappa\cdot(\vU\cdot\nabla )\vU_Q   
 +\vU_\kappa\cdot(\vB\cdot\nabla )\vB_Q -\vB_\kappa\cdot(\vU\cdot\nabla )\vB_Q   
 +\vB_\kappa\cdot(\vB\cdot\nabla )\vU_Q +\epsilon\vJ_\kappa\cdot(\vB\times\vJ_Q) \Big]
\label{Pi} 
\end{equation}
\end{widetext}

\begin{figure}
\includegraphics{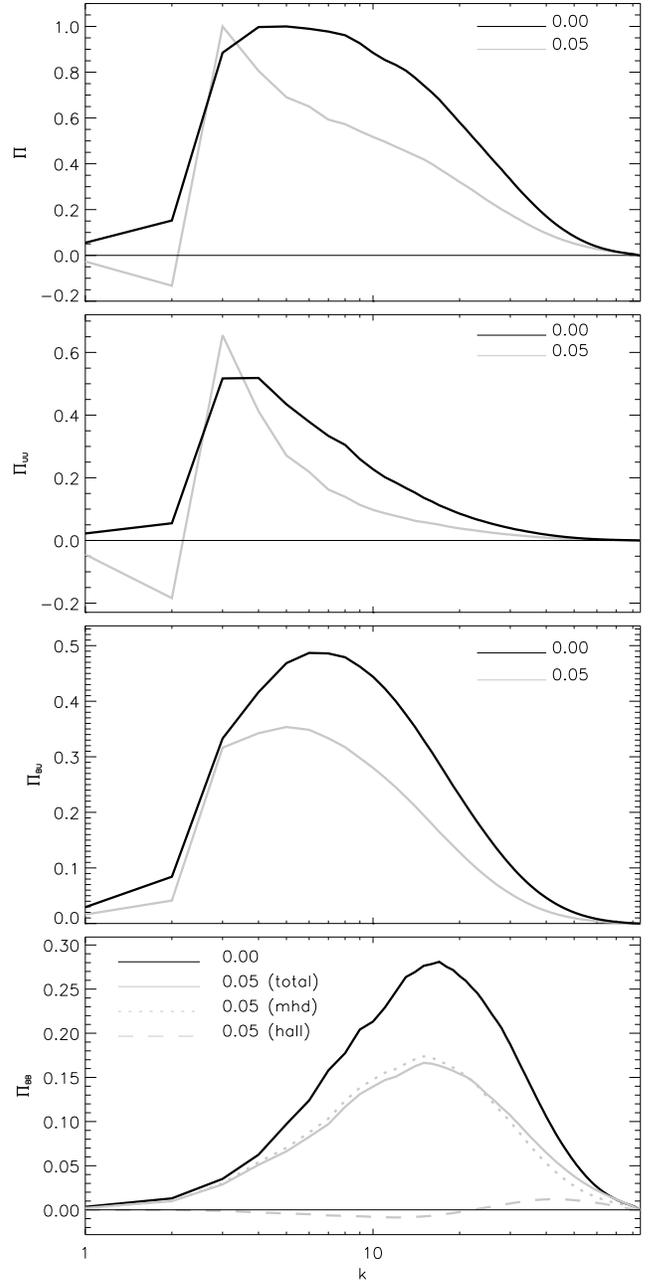}
\caption{Energy fluxes vs.~$k$, normalized by the total dissipation rate D. The black line corresponds to a time average at the stationary regime of the run with $\epsilon=0$, while the gray line is for $\epsilon=0.05$. }\label{fig:flux}
\end{figure}

Each of the five terms on the right-hand side of Eq.~(\ref{Pi}) has a straightforward interpretation. The first term (hereafter called $\Pi_{UU}$) is a purely kinetic energy flux, which is already present in hydrodynamic turbulence, and is responsible of the direct energy cascade in that particular case. The second and fourth terms add to zero (i.e., $\Pi_{UB} + \Pi_{BU} =0$), since they correspond to the exchange between kinetic and magnetic energy by Lorentz forces and Faraday induction. The third term ($\Pi_{BB}$) is flux of  magnetic energy associated to the advection of magnetic field lines by the velocity field, already present in the MHD case. Finally, the fifth term ($\Pi_{BB}^{Hall}$) is also a flux of magnetic energy, but exclusively due to the nonlinearity introduced by the Hall current. The first frame of Fig.~\ref{fig:flux} shows the total energy flux vs.~$k$ for the runs with $\epsilon=0$ (black line) and $\epsilon=0.05$ (gray line). The next three frames disaggregate the energy flux into the various parts listed above.

The energy fluxes displayed in Fig. \ref{fig:flux} confirm the observation made when considering the shell-to-shell transfer functions. Note that these energy fluxes have been normalized by the (time averaged) total dissipation rate to allow a comparison between runs with different Hall parameter values. The flux associated to the Hall term slows down the cascade of magnetic energy toward small scales for wavenumbers smaller than the Hall wavenumber, since $\Pi_{BB}^{Hall}$ is negative in this range. At the same time, the Hall effect speeds up the energy cascade at smaller scales, where the $\Pi_{BB}^{Hall}$ becomes positive. Note that the change of sign takes place at the Hall scale (see the dashed line in the lowest panel of Fig. \ref{fig:flux}). This result explains why the current spectrum was observed to peak near the Hall scale, and the associated decrease in the magnetic energy dissipation rate as the amplitude of the Hall term was increased. The slow-down in the transfer of magnetic energy towards small scales (where it finally dissipates) is also responsible for the faster build up of magnetic
energy through dynamo action in the Hall-MHD case.

Note that although the Hall transfer rate is three orders of magnitude smaller than the total energy transfer rate (see Fig.~\ref{fig:trans}), the contribution of the Hall term to the flux is not negligible. This is the result of the strong non-locality of the Hall term: while the Hall transfer is small for each value of $\kappa$ and $Q$, the flux at the wavenumber $k$ results from summing over all values of $Q$, and over all values of $\kappa < k$. The slow decay of the Hall transfer far from the diagonal $\kappa=Q$ in Fig.~\ref{fig:trans} (associated to the non-locality) gives a substantial contribution to the total flux.

The other fluxes are also modified by the Hall term. The total (i.e., $\Pi$) and $\Pi_{UU}$ fluxes decay faster with increasing wavenumber in the Hall-MHD case, and show the build up of (mechanical) flows at scales larger than the forcing scale (indicated by the negative value of $\Pi_{UU}$ for small wavenumbers). Such an effect for Hall-MHD has been predicted from theoretical models \cite{Mahajan98} and confirmed by numerical simulations \cite{Mininni05}. Also, the $\Pi_{UB}$ flux peaks at smaller wavenumbers in the Hall-MHD case.

The modification of the fluxes is consistent with the changes in the global dissipation rates. As the Hall term reduces the transfer of magnetic energy to smaller scales at scales larger than the Hall scale, the global dissipation of magnetic energy decreases. Note that this result is compatible with studies \cite{Birn01,Smith04}, which find faster reconnection rates in Hall-MHD simulations with current sheets initially set up at small scales. For spatial structures such as current sheets, at scales smaller than the Hall scale, the Hall term increases the transfer toward even smaller scales. As a result, the dissipation rate at those particular scales is also increased, which is opposite to the result obtained for the global dissipation rate.

\section{Large magnetic Prandtl number}\label{sec:Pm}

We have so far considered the particular case $Pm = 1$, i.e., $\eta = \nu$. However, there are several low-density and high-temperature astrophysical plasmas which are characterized by $Pm = \nu /\eta \gg 1$, such as the interstellar medium, intracluster gas in between of galactic clusters, jets, or accretion disks. We performed numerical simulations with $Pm = 10$, so that the viscous dissipation wavenumber $k_\nu$ becomes much smaller than the resistive dissipation wavenumber $k_\eta$. As a result, magnetic fluctuations in this large-$Pm$ regime may grow at subviscous scales.

\begin{figure}
\includegraphics{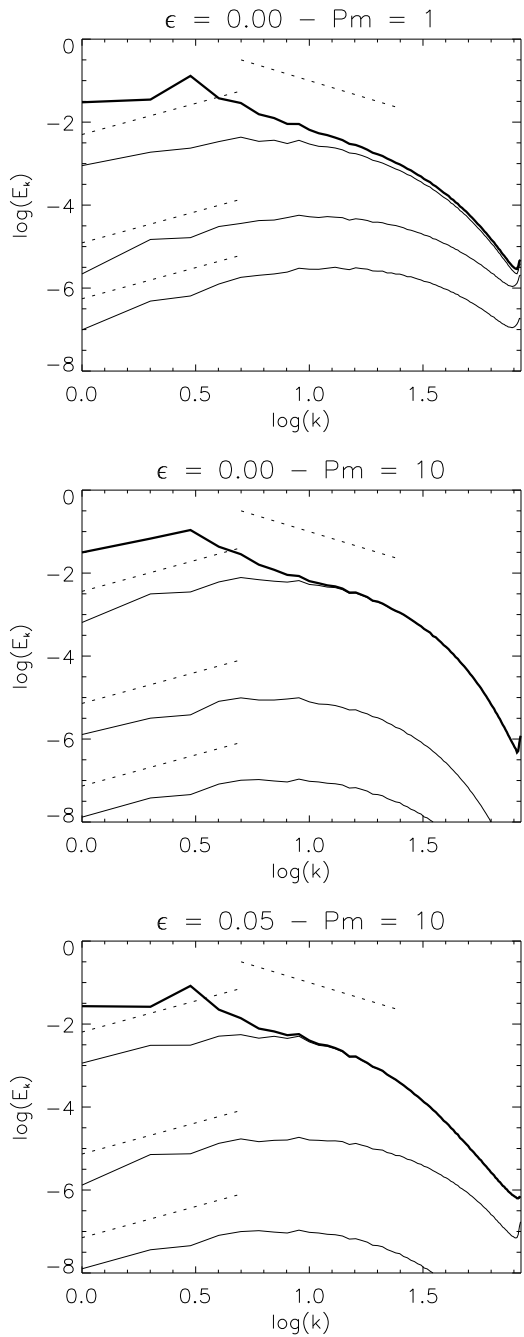}
\caption{Total energy spectrum (thick trace) at $t=72$ for three different runs (as labeled). Magnetic energy spectra at $t = 18, 36, 72$ (thin lines from bottom to top in each panel) are also shown. The Kolmogorov and Kazantsev slopès are overlaid (dotted trace) for reference.}\label{fig:sppm}
\end{figure}
\begin{figure}
\includegraphics{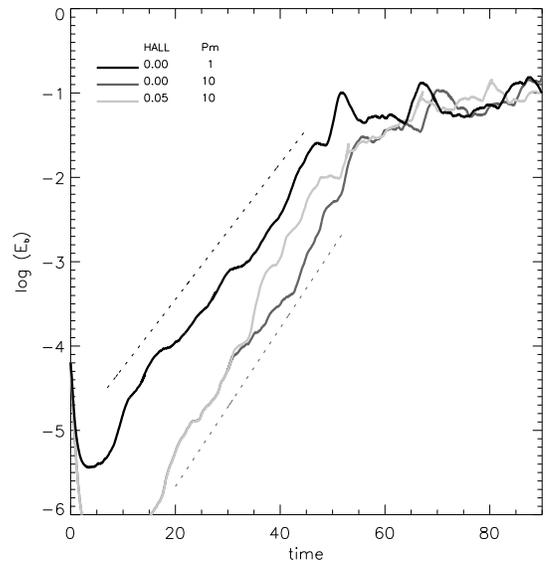}
\caption{Magnetic energy vs.~time for the three runs as labeled.}\label{fig:b2pm}
\end{figure}

\begin{table}
\caption{\label{tab:2} Global results for runs with different values of the magnetic Prandl number $Pm$; $\epsilon$ is the amplitude of the Hall effect, $E_b /E$ is the ratio of magnetic to total energy, $k_J$ is the magnetic Taylor  wavenumber for the current density distribution, and $D_b /D$ is the ratio of magnetic to total dissipation. }
\begin{ruledtabular}
\begin{tabular}{ccccccccc}
 & $\epsilon$ & $Pm$ & E & $E_b /E $ & $k_J$ & D & $D_b /D $ & \\
\hline
 & 0.00 &  1 & 0.37 & 0.14 & 23.1 & 0.13 & 0.48 & \\
 & 0.00 & 10 & 0.29 & 0.25 & 16.4 & 0.14 & 0.37 & \\
 & 0.05 & 10 & 0.26 & 0.19 & 14.3 & 0.11 & 0.21 & \\
\end{tabular}
\end{ruledtabular}
\end{table}

In Figure \ref{fig:sppm} we show the energy spectra for three different runs (labeled). In the top frame we repeat the spectra from the purely MHD run with $Pm = 1$ as a reference. In the central frame we show the spectra for an MHD run (i.e., $\epsilon = 0$), but with $Pm = 10$. The thick trace corresponds to the total energy spectrum at $t = 72$, corresponding to the saturation of the dynamo. The various thin curves correspond to the magnetic energy spectrum at the successive times $ t = 18$, $36$, and $72$ from bottom to top. In the large-$Pm$ regime, the magnetic field grows in the sub-viscous region of the spectrum. It is apparent that for $Pm = 10$ most of the energy at small scales (i.e., large $k$'s) is magnetic. These results are consistent with those reported in Ref.~\cite{Haugen04} from very similar non-helical simulations. The Kazantsev spectrum $E_k \approx k^{3/2}$ is also a good approximation at large scales, although at large-$Pm$ is less surprising, since the kinetic energy is more confined toward the small-$k$ spectral region.

The lower panel in Fig. \ref{fig:sppm} corresponds to a simulation with moderate Hall value ($\epsilon = 0.05$) and large magnetic Prandtl number ($Pm = 10$). By comparison with the case displayed in the central panel ($\epsilon = 0$ and $Pm = 10$), the dynamo efficiency is larger in the presence of the Hall effect, as also confirmed by Fig. \ref{fig:b2pm}.

Figure \ref{fig:b2pm} shows the growth of magnetic energy in the three simulations in lin-log scale. In the MHD case, the linear growth rate at large-$Pm$ (for the same magnetic diffusivity $\eta$) is somewhat larger than in the $Pm=1$ case, which can be expected as the flow is smoother in the former case and turbulent magnetic diffusion is therefore less effective. In the Hall-MHD case, the linear regime is again found to be followed by a non-linear stage when the Hall-effect becomes relevant and the magnetic field is advected by the electron velocity, as found in the simulations with $Pm=1$.

\begin{figure}
 \includegraphics{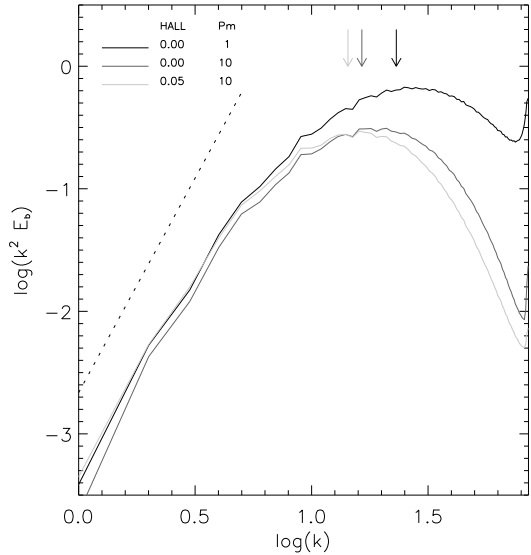}
\caption{Spectral distribution of current density, i.e. $k^2 E_b(k)$ vs.~$k$ for three runs with different values of $\epsilon$ and $Pm$ (labeled). The dotted trace correspond to the Kazantsev slope $k^{7/2}$. The arrows indicate the average wavenumber $k_J$ (see Eqn.~\ref{kJ}) for each of the current density distributions.}\label{fig:j2kpm}
\end{figure}

Other features of the Hall-MHD small-scale dynamos reported in the $Pm=1$ simulations can also be identified in the $Pm=10$ case. As examples, Table \ref{tab:2} shows saturation values of the total energy, total dissipation, and the ratios of magnetic to total energy and magnetic to total dissipation for the runs in Figs.~\ref{fig:sppm} and \ref{fig:b2pm}. In the MHD case, the increase of the magnetic Prandtl number moves the peak of the current density spectrum towards smaller wavenumbers (see the values of $k_J$ in Table \ref{tab:2} and Fig. \ref{fig:j2kpm}). As discussed in Sect.~\ref{sec:spec}, the Hall effect moves this peak further to larger scales.

\section{Conclusions}\label{sec:conclu}

We present results from three dimensional simulations of small-scale dynamo action for magnetic Prandtl numbers $Pm=1$ and $10$ in conducting flows with the Hall effect. This effect is believed to be non-negligible in sufficiently diffuse media, and its relevance has been recognized in various astrophysical, space, and laboratory plasmas. As a first step toward a better description of dynamo action in such media, only the incompressible Hall-MHD equations were solved, and the inclusion of compressible effects as well as other kinetic effects such as ambipolar difussion is left for future studies.

However, the inclusion of only the Hall effect acting at the smallest relevant dynamical scales of the flow gives rise to measurable differences with previous studies of dynamo action. A magnetic non-linear regime is identified when the magnetic field (and the current density) becomes large enough to differentiate the electron velocity from the bulk flow velocity. After saturation, differences in the stationary level of magnetic energy and in the total and magnetic energy dissipation rates are obtained, depending on the amplitude of the Hall effect. Finally, the peak of the current density spectrum is found to be dependent on the strength of the Hall term, with its peak moving toward larger scales (smaller wavenumbers) as the Hall scale is increased.

By studying the detailed transfer of energy among fields and scales, we observe that the effect of the Hall term is twofold: it transfers energy towards larger scales for scales larger than the Hall length, and it transfers energy towards smaller scales for scales smaller than this length. The modification of the energy flux resulting from this transfer is consistent with the observed changes in the saturation values of energy and dissipation rate observed in our simulations.

\begin{acknowledgments}
The authors acknowledge support from UBACYT grants X468/08, X469/08, and X092/08, and from PICT grants 2005-33370, 2007-02211 and 2007-00856.
\end{acknowledgments}

\end{document}